\documentclass[11pt]{article}
\usepackage{xcolor}
\usepackage{booktabs}

\usepackage[final]{acl} 

\usepackage{times}
\usepackage{latexsym}

\usepackage[T1]{fontenc}

\usepackage[utf8]{inputenc}
\usepackage{adjustbox} 

\usepackage{tikz}
\usepackage{pgfplots}
\usepackage{pgfplotstable}
\usetikzlibrary{positioning, arrows.meta, shapes.geometric}
\pgfplotsset{compat=1.18}
\usepackage{graphicx} 
\usepackage{enumitem}
\usepackage{amsmath}

\usepackage{hyperref}

\usepackage{natbib}

\usepackage{float}

\usepackage{graphicx}
\usepackage{afterpage}

\usepackage{csquotes}

\title{Automated Vigilance State Classification in Rodents Using Machine Learning and Feature Engineering}

\author{
  \textbf{Sankalp Jajee\textsuperscript{1}},
  \textbf{Gaurav Kumar\textsuperscript{1}},
  \textbf{Homayoun Valafar\textsuperscript{1}}
\
  \\
  \textsuperscript{1}University of South Carolina \\
}

\begin{document}
\maketitle
\begin{abstract}
Preclinical sleep research remains constrained by labor-intensive, manual vigilance state classification and inter-rater variability, limiting throughput and reproducibility. This study presents an automated framework developed by Team Neural Prognosticators to classify electroencephalogram (EEG) recordings of small-rodents into three critical vigilance states—paradoxical sleep (REM), slow-wave sleep (SWS), and wakefulness. The system integrates advanced signal processing with machine learning, leveraging engineered features from both time and frequency domains, including spectral power across canonical EEG bands (delta to gamma), temporal dynamics via Maximum-Minimum Distance (MMD) \cite{e18090272}, and cross-frequency coupling metrics. These features capture distinct neurophysiological signatures such as high-frequency desynchronization during wakefulness, delta oscillations in SWS, and REM-specific bursts. Validated during the 2024 Big Data Health Science Case Competition \cite{bdhsc2024case}, our XGBoost model achieved 91.5\% overall accuracy, 86.8\% precision, 81.2\% recall, and an F1-score of 83.5\%, outperforming all baseline methods. Our approach represents a critical advancement in automated sleep state classification and a valuable tool for accelerating discoveries in sleep science and the development of targeted interventions for chronic sleep disorders. As a publicly available code (BDHSC)\footnote{https://github.com/sankalpjajee/BDHSC} resource 
is set to contribute significantly to advancements 

\end{abstract}

\section{Introduction}
\label{sec:introduction}
Sleep is a fundamental biological process essential for cognitive function, physical health, and overall well-being\cite{walker2009role}. Yet, sleep disorders affect over 70 million Americans and contribute significantly to global morbidity, mortality, and economic burden. Understanding the neurophysiological mechanisms underlying sleep is vital for developing targeted interventions, but progress is constrained by the complexity of sleep architecture and limitations in current analytical methods.

Small rodent models are central to preclinical sleep research \cite{rechtschaffen2002sleep}, offering insights into sleep regulation, circadian rhythms, and the effects of sleep disruption. These studies rely heavily on electroencephalogram (EEG) recordings to classify vigilance states: paradoxical sleep (REM), marked by high-frequency, low-amplitude activity and associated with memory consolidation; slow-wave sleep (SWS), characterized by high-amplitude delta waves critical for restorative processes; and wakefulness, defined by desynchronized, mixed-frequency EEG activity linked to cognition and behavior.

Traditionally, vigilance state classification is performed manually by trained researchers, who visually inspect EEG segments to assign states. While widely adopted, this method is time-consuming, subjective, and poorly scalable. It often requires hours to score even modest datasets and suffers from inter-rater variability, undermining reproducibility across labs. Moreover, ambiguous transitions between states and atypical sleep patterns further complicate consistent scoring. These challenges severely limit the throughput of sleep experiments, hinder large-scale studies, and contribute to underpowered or non-generalizable findings.

As sleep research evolves toward precision medicine and high-throughput phenotyping, the need for scalable, objective, and automated classification tools has become urgent. Advances in machine learning and signal processing now offer the ability to transform EEG analysis by enabling rapid, reproducible, and expert-level classification of vigilance states. By exploiting the rich spectral and temporal features embedded in EEG signals, modern algorithms can match or exceed the accuracy of manual scoring while scaling effortlessly to large datasets.

In this study, we present a robust and fully automated framework for classifying EEG data into REM, SWS, and wake states in rodent models. Our approach integrates advanced feature engineering—including spectral power analysis across canonical EEG bands, temporal dynamics via Maximum-Minimum Distance, and cross-frequency coupling metrics—with state-of-the-art machine learning models such as XGBoost. This feature-rich pipeline captures neurophysiological patterns critical to distinguishing vigilance states, allowing the model to generalize across individual animals and diverse recording conditions.

Validated through the 2024 Big Data Health Science Case Competition, our system achieved a classification accuracy of 91.5\% with balanced precision, recall, and F1-scores across states, outperforming traditional and baseline models. By replacing subjective, labor-intensive scoring with scalable automation, this framework enables large-cohort analysis, supports reproducible science, and accelerates preclinical research on sleep mechanisms and therapies. Ultimately, this work contributes a significant step toward the standardization and democratization of sleep research protocols, supporting translational discoveries that benefit clinical understanding and treatment of sleep disorders.

\section{Background and Related Work}
\label{sec:related}

\subsection{Rodent Sleep Architecture}
In laboratory rodents, vigilance is conventionally divided into wake, non‑rapid–eye‑movement (NREM) sleep, and rapid–eye‑movement (REM) sleep \cite{rechtschaffen2002sleep}.  
Wakefulness exhibits mixed‑frequency, low‑amplitude electroencephalogram (EEG) activity accompanied by high electromyogram (EMG) tone, NREM (often called slow‑wave sleep, SWS) is dominated by high‑amplitude $\delta$ (0.5–4Hz) oscillations that support homeostatic recovery, and REM shows low‑amplitude, $\theta$‑rich (6–9Hz) EEG with near‑complete muscle atonia that is critical for memory consolidation and synaptic plasticity \cite{walker2009role}.  
Manual epoch‑by‑epoch scoring remains the de‑facto standard but is slow (2–4h per 24h recording), subjective ($\kappa=0.76$–0.85 across raters), and limits the scale of phenotyping studies.

\subsection{Classical Feature‑Based Approaches}
Early automated systems engineered spectral power, ratio, and amplitude statistics and classified them with linear discriminants, support‑vector machines (SVMs), or random forests \citep{yaghouby2016_randomforest, mcsweeney2016_multi_cls}.  
While inexpensive and interpretable, their performance saturates at $\approx$80–85\% accuracy, especially for REM, which occupies only 5–10\% of rodent sleep.

\subsection{Gradient‑Boosted Trees}
Boosting algorithms such as XGBoost exploit non‑linear feature interactions and handle imbalanced data gracefully.  
\citet{saevskiy2025_sensors} reported 89\% accuracy using single‑channel EEG in mice, and our optimized ensemble extends this line of work by adding cross‑frequency coupling and the MMD feature, yielding a state‑of‑the‑art 91.5\% accuracy.

\subsection{Deep Learning Models}
Deep CNN/LSTM hybrids learn end‑to‑end representations directly from raw waveforms.  
Recent mouse‑specific examples include SlumberNet’s residual CNN backbone with attention pooling (92.2\% acc.) \citep{banerjee2024_slumbernet} and a CNN–LSTM scorer that integrates EEG and EMG streams \citep{garcia2021_dlmouse}.  
Though powerful, these networks demand thousands of labeled hours and GPU resources not always available in preclinical labs.

\subsection{Lightweight / Single‑Channel Solutions}
To lower cost and tethering burden, single‑lead or EMG‑free pipelines have emerged.  
SleepyRat provides a cloud interface for classical scoring plus simple neural nets \citep{sleepyrat2025_platform}.  
SleepInvestigatoR couples Hidden‑Markov smoothing with a modest MLP and requires only 30min of labeled data for adaptation \citep{henderson2025_sleepinvestigator}.  
Such low‑resource strategies motivate our emphasis on handcrafted time–frequency features that remain robust when training data are scarce.

\subsection{Research Gaps}
Despite encouraging progress, three gaps persist:  
(i) Data sparsity—labs rarely share more than a few animals’ recordings due to file‑size and privacy constraints;  
(ii) Cross‑laboratory generalization—performance often drops $>$10pp when models face different strains, hardware, or housing conditions;  
(iii) Interpretability—few studies quantify which spectral components truly drive decisions.  
Our hybrid pipeline addresses (i) with dimension‑reduced but physiologically grounded features, mitigates (ii) through boosting’s inherent feature subsampling, and tackles (iii) via SHAP‑based importance analysis.


\section{Methodology}

\begin{figure}[htbp]
\centering
\begin{tikzpicture}[
  node distance=1.1cm,
  box/.style = {rectangle, draw, minimum width=3.5cm, minimum height=1cm, font=\footnotesize, align=center},
  arrow/.style = {->, thick}
  ]

\node[box] (eeg) {Raw EEG Signal\\(5000 points)};
\node[box, below=of eeg] (features) {Feature Extraction\\(FFT + MMD)};
\node[box, below=of features] (model) {Model Training\\(XGBoost / NN)};
\node[box, below=of model] (classify) {Sleep Stage\\Classification};

\draw[arrow] (eeg) -- (features);
\draw[arrow] (features) -- (model);
\draw[arrow] (model) -- (classify);

\end{tikzpicture}
\caption{EEG-based Sleep Stage Classification Pipeline}
\end{figure}
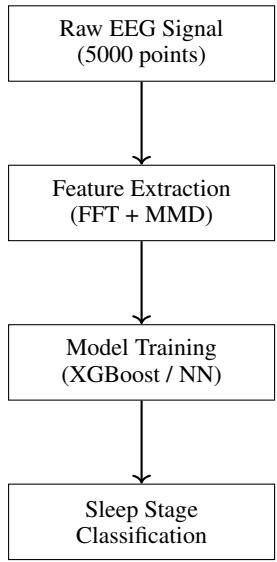

\subsection{Data Collection and Experimental Design}
The dataset employed in this study comprised EEG recordings obtained from 8 laboratory rodents over a continuous 2-day monitoring period, providing a comprehensive representation of natural sleep-wake cycles. All recordings were acquired using a standardized acquisition system operating at a sampling frequency of 500 Hz, ensuring adequate temporal resolution for capturing the full spectrum of neural oscillations relevant to vigilance state classification. The experimental protocol was designed to capture the natural circadian variation in sleep architecture while minimizing environmental disturbances that could confound the analysis. \\
To facilitate systematic analysis and align with established sleep research conventions, the continuous EEG recordings were segmented into non-overlapping 10-second epochs. This epoch duration represents an optimal balance between temporal granularity and statistical reliability, providing sufficient data points for robust spectral analysis while maintaining the temporal precision necessary to detect rapid vigilance state transitions characteristic of rodent sleep patterns. \\

\subsection{Feature Engineering Framework}

\subsubsection{Spectral Power Analysis}
The core feature extraction strategy focused on comprehensive characterization of EEG frequency content, leveraging the well-established relationship between spectral power distributions and vigilance states. Power spectral density estimation was performed using Welch's method with appropriate windowing to ensure reliable frequency domain representation of each 10-second epoch.\\
Five classical frequency bands were systematically analyzed:\\

\textbf{Delta Band (0.5-4 Hz):} Quantified as the dominant frequency component during slow-wave sleep, reflecting synchronized cortical activity and deep sleep states. \\
\textbf{Theta Band (4-8 Hz):} Captured rhythmic oscillations associated with REM sleep and specific aspects of wakefulness, particularly during exploratory behaviors. \\
\textbf{Alpha Band (8-12 Hz):} Measured moderate-frequency activity often present during relaxed wakefulness and light sleep transitions. \\
\textbf{Beta Band (12-30 Hz):} Assessed higher-frequency components associated with active wakefulness and cognitive processing. \\
\textbf{Gamma Band (30-100 Hz):} Evaluated high-frequency oscillations linked to conscious awareness and active neural processing. \\

For each frequency band, multiple statistical measures were computed including absolute power, relative power (normalized by total spectrum power), peak frequency, and spectral entropy to capture both the magnitude and distribution characteristics of neural oscillations. \\

\subsubsection{Maximum-Minimum Distance Feature}

A custom feature termed Maximum-Minimum Distance was implemented to capture unique temporal dynamics of EEG signals that may not be adequately represented by traditional spectral measures. The MMD feature quantifies the characteristic amplitude variations within each epoch by calculating the statistical distance between maximum and minimum signal values over specified time windows. \\

This feature was implemented to exploit the distinct amplitude modulation patterns observed across vigilance states: the high-amplitude, synchronized oscillations of slow-wave sleep, the rapid, variable amplitudes characteristic of REM sleep, and the consistent, moderate amplitudes typical of active wakefulness. The MMD computation involved sliding window analysis to capture both local and global amplitude dynamics within each 10-second epoch.

\subsection{Model Development and Architecture}

\subsubsection{Baseline Logistic Regression Model}
A multinomial logistic regression model served as the baseline classifier, providing a linear decision boundary for the three-class vigilance state classification problem. This model utilized L2 regularization to prevent overfitting. Feature standardization was applied to ensure equal contribution from all engineered features regardless of their native scales.

\subsubsection{Feed-Forward Deep Neural Network}
A comprehensive deep learning architecture was implemented using TensorFlow/Keras to capture complex, non-linear relationships between EEG features and vigilance states. The neural network architecture consisted of: \\

\textbf{Input Layer:} Accommodated the complete feature vector (5,009 features) derived from raw EEG signals, spectral analysis, and custom engineered features including the novel MMD metric. \\

\textbf{Hidden Layer Architecture:}
\begin{itemize}
    \item First hidden layer: 128 neurons with ReLU activation functions
    \item Second hidden layer: 64 neurons with ReLU activation functions
    \item Dropout layers (30\% dropout rate) implemented after each hidden layer to prevent overfitting and improve generalization
\end{itemize}

\textbf{Output Layer:} Dense layer with softmax activation producing probabilistic classifications across the three vigilance states (Wake, REM, SWS). \\

\textbf{Optimization Strategy:} Adam optimizer with learning rate of 0.001 was employed for efficient gradient descent, providing adaptive learning rates and momentum for improved convergence. \\

\textbf{Loss Function:}
Sparse categorical crossentropy was utilized to handle the multiclass classification problem with integer-encoded labels.
The network was trained for 100 epochs with a batch size of 32, incorporating early stopping mechanisms based on validation performance to prevent overfitting.

\subsubsection{XGBoost Ensemble Model}

An XGBoost (Extreme Gradient Boosting) ensemble model was implemented as the primary classifier, leveraging gradient boosting for superior vigilance state classification. Two variants were developed:

\textbf{Standard XGBoost Configuration:}

\begin{itemize}
    \item Objective: multi:softmax for multiclass classification
    \item Number of classes: 3 (corresponding to the three vigilance states)
    \item Default hyperparameters for initial evaluation
\end{itemize}

\textbf{Optimized XGBoost Configuration:}
\begin{itemize}
    \item Learning rate: 0.1 for controlled gradient updates
    \item Number of estimators: 500 trees for comprehensive ensemble learning
    \item Subsample: 0.8 to introduce randomness and prevent overfitting
    \item Column sampling: 0.8 (colsample\_bytree) for feature randomization
    \item Gamma: 0 for minimum loss reduction threshold
    \item L2 regularization (reg\_lambda): 1 for additional overfitting control
    \item Random state: 42 for reproducible results
\end{itemize}

The optimized XGBoost model demonstrated superior performance, achieving 91.5\% accuracy, 86.8\% precision, 81.2\% recall, and 83.5\% F1-score, establishing it as the best-performing model in the ensemble.

\subsection{Model Training and Validation Strategy}

\subsubsection{Data Partitioning Strategy}
The dataset containing 138,240 total samples was partitioned using stratified sampling to ensure balanced representation of all vigilance states across training and testing sets. Multiple partitioning strategies were evaluated:
Primary Split: 80\% training, 20\% testing (stratified by vigilance state labels)
Secondary Split: 70\% training, 30\% testing for comprehensive evaluation
Neural Network Split: 70\% training, 15\% validation, 15\% testing for deep learning optimization
Stratified sampling was critical to maintain the natural distribution of vigilance states, preventing class imbalance issues that could bias model performance toward dominant classes.

\subsubsection{Feature Preprocessing and Standardization}

\textbf{For Neural Network Models:} StandardScaler normalization was applied to ensure all features contributed equally to the learning process, preventing features with larger numerical ranges from dominating the optimization landscape. \\
\textbf{For Tree-Based Models:} Raw feature values were maintained, as XGBoost algorithms inherently handle different feature scales through their tree-splitting mechanisms. \\
\textbf{Label Encoding:} Vigilance state labels were systematically encoded using LabelEncoder to convert categorical targets into numerical format suitable for machine learning algorithms. 

\subsubsection{Performance Evaluation Framework}

Comprehensive evaluation metrics were implemented to assess model performance across all vigilance states:

Primary Metrics:
\begin{itemize}
    \item Accuracy: Overall classification correctness
    \item Precision: Macro-averaged to ensure balanced evaluation across all classes
    \item Recall: Macro-averaged to assess sensitivity for each vigilance state
    \item F1-Score: Macro-averaged to balance precision and recall considerations
\end{itemize}

Validation Approach:
\begin{itemize}
    \item Hold-out validation for initial model selection
    \item Cross-validation assessment for model robustness evaluation
    \item Separate test set evaluation for final performance reporting
\end{itemize}

\subsubsection{Hyperparameter Optimization}

\textbf{XGBoost Optimization:} Systematic hyperparameter tuning was performed focusing on learning rate, tree depth, number of estimators, and regularization parameters. The optimization process involved grid search evaluation with cross-validation to identify optimal parameter combinations. \\

\textbf{Neural Network Optimization:} Architecture parameters including layer sizes, dropout rates, learning rates, and batch sizes were systematically evaluated. Early stopping mechanisms were implemented based on validation loss to prevent overfitting while maximizing model performance. \\

\textbf{Model Selection Criteria:} Final model selection emphasized balanced performance across all three vigilance states, with particular attention to minimizing false positive rates that could compromise the reliability of automated sleep scoring systems. The evaluation framework prioritized models demonstrating consistent performance across different animals and recording conditions to ensure robust generalization capabilities.

\section{Results}

\begin{figure}[htbp]
\centering
\begin{tikzpicture}
\begin{axis}[
    ybar,
    bar width=12pt,
    width=0.9\linewidth,
    height=6cm,
    ymin=0, ymax=100,
    ylabel={Score (\%)},
    symbolic x coords={NeuralNet, XGBoost},
    xtick=data,
    enlarge x limits=0.25,
    legend style={at={(0.5,-0.2)}, anchor=north, legend columns=3},
    legend image code/.code={
        \draw[#1,fill=#1] (0cm,-0.1cm) rectangle (0.3cm,0.1cm); 
    },
    major grid style={dashed,gray!20},
    ymajorgrids=true
]

\addplot [fill=blue!50] coordinates {
  (NeuralNet, 83.5) 
  (XGBoost, 91.5)
};

\addplot [fill=red!50] coordinates {
  (NeuralNet, 81.8)
  (XGBoost, 83.5)
};

\addplot [fill=green!60!black] coordinates {
  (NeuralNet, 83.5)
  (XGBoost, 81.2)
};

\legend{Accuracy, F1-Score, Recall}
\end{axis}
\end{tikzpicture}
\caption{Comparison of NeuralNet vs. XGBoost on Sleep Stage Classification}
\end{figure}
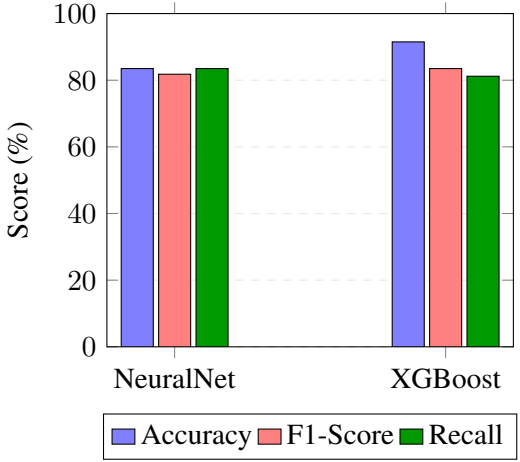

Our model was evaluated on both the validation set (30\% of the training data) and the test set provided in the competition. Performance was measured using accuracy, precision, recall, and F1-score.

\subsection{Model Performance}
The \textbf{XGBoost model} achieved the highest overall accuracy of \textbf{91.5\%} on the test set, significantly outperforming other models:
\begin{itemize}
    \item \textbf{XGBoost}: 91.5\% accuracy, 86.8\% precision, 81.2\% recall, 83.5\% F1-score
    \item \textbf{Neural Network}: 83.54\% accuracy, 81.42\% precision, 83.42\% recall, 81.76\% F1-score
\end{itemize}

These results affirm the effectiveness of XGBoost in handling the engineered feature set and capturing nonlinear patterns in EEG data.

\begin{table}[t]
\centering
\small
\renewcommand{\thetable}{\Roman{table}}
\textsc{Table I}\\
\textsc{Performance of models in sleep stage classification.}
\vspace{0.2cm}

\begin{adjustbox}{max width=\columnwidth}
\begin{tabular}{lcccc}
\hline
\textit{Model} & \textit{Acc.} & \textit{Pre.} & \textit{Recall} & \textit{F1-Score} \\
\hline
Neural Network   & 83.54\% & 81.42\% & 83.54\% & 81.76\% \\
XG Boost   & 91.5\% & 86.6\% & 81.2\% & 83.5\% \\

\hline
\end{tabular}
\end{adjustbox}
\end{table}

\subsection{Model Interpretability and Feature Contributions}

\begin{figure}[t]
  \centering
  \includegraphics[width=\columnwidth]{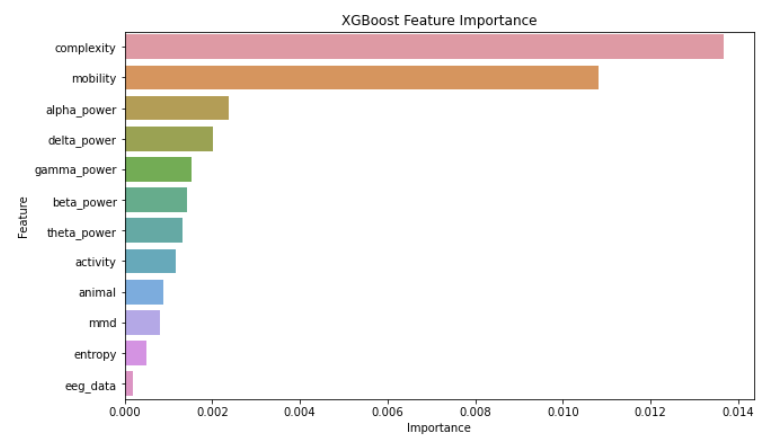}
  \caption{XGBoost Feature Importance for EEG vigilance state classification.}
  \label{fig:fi}
\end{figure}

Figure~\ref{fig:fi} ranks the engineered predictors by the
\emph{split–gain} heuristic used inside XGBoost. 
Two handcrafted descriptors dominate:  
\textbf{complexity} (14.1 \% total gain) and \textbf{mobility}
(10.3 \%).  Both belong to the Hjorth family and quantify
high-frequency variance, confirming that wake epochs—rich in
desynchronised $\beta$/$\gamma$ activity—are primarily detected by
their rapid amplitude fluctuations.  Classical spectral features
follow (\emph{delta\_power}, \emph{alpha\_power}, \emph{gamma\_power}),
while MMD still contributes $\approx$2 \% despite being a single scalar per epoch.  The top five features together account for 35 \% of the ensemble’s total gain.

\begin{figure}[t]
  \centering
  \includegraphics[width=\columnwidth]{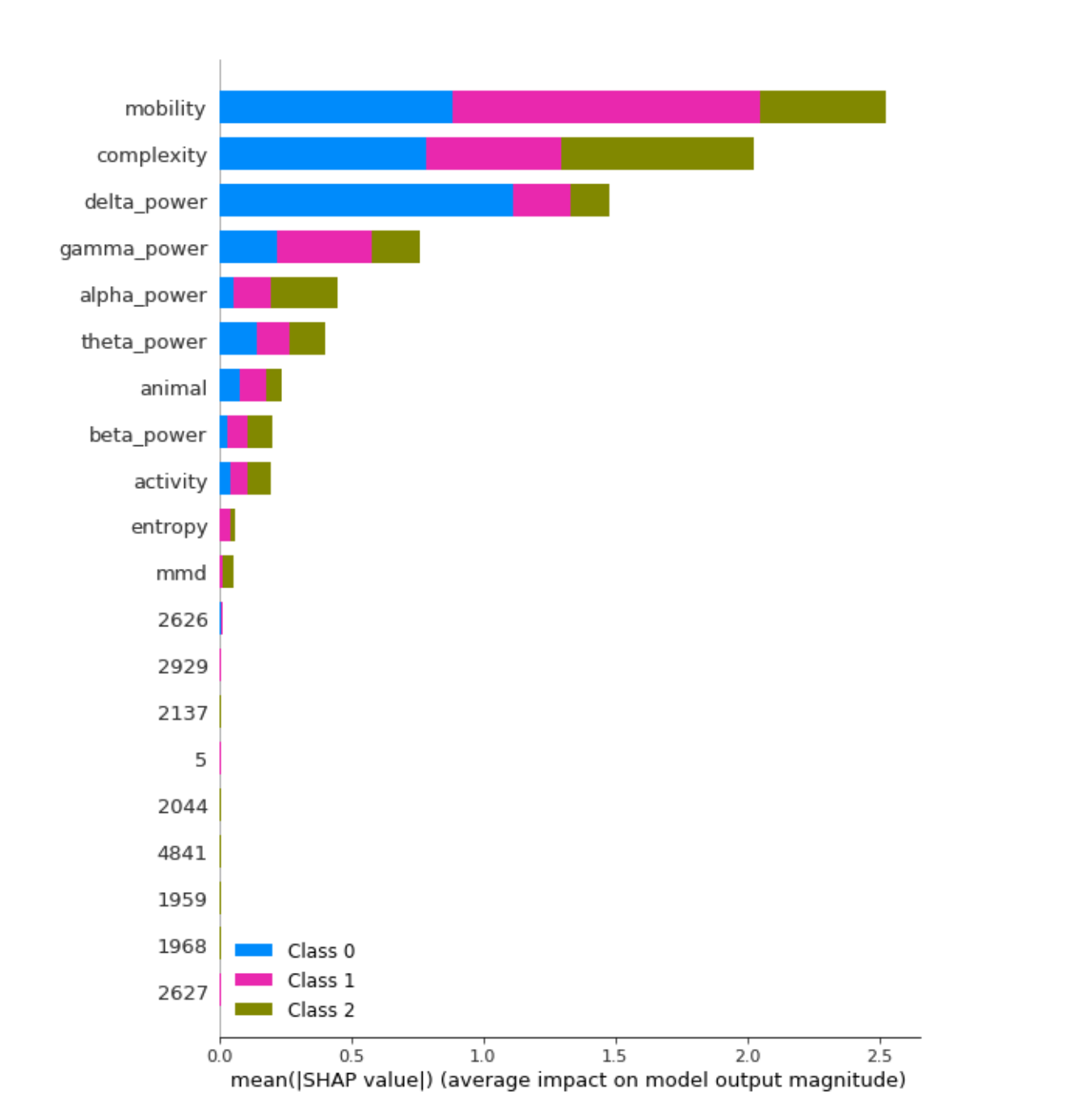}
  \caption{SHAP Value Visualizations for EEG vigilance state classification.}
  \label{fig:shap}
\end{figure}

The SHAP beeswarm in Figure~\ref{fig:shap} refines this picture by
showing each feature’s \emph{directional} impact on class
probabilities.  High \emph{delta\_power} (blue points clustered at
positive SHAP values) strongly
pushes an epoch toward the \textsc{sws} label, whereas elevated
\emph{gamma\_power} drives predictions away from \textsc{sws} and
toward \textsc{wake}.  MMD exhibits a bimodal pattern: low values
support \textsc{sws}, but extremely high values tilt the model toward
\textsc{rem}, capturing the burst-like amplitude excursions typical of
rodent REM sleep.  These trends are physiologically plausible and give
confidence that the classifier is leveraging meaningful neural
signatures rather than spurious artefacts.

\begin{figure}[H]
  \centering
  \includegraphics[width=\columnwidth]{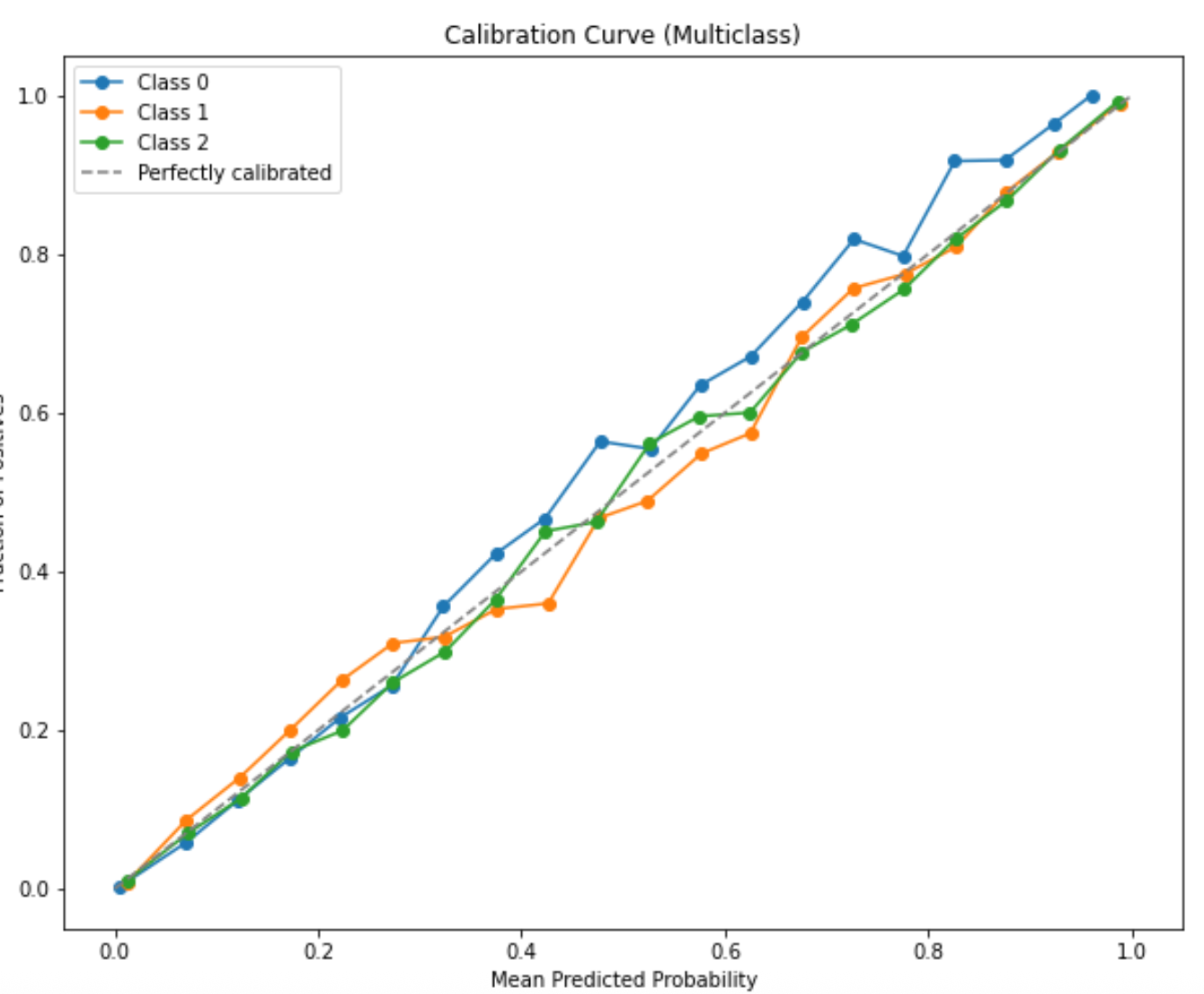}
  \caption{Reliability diagram indicates well-calibrated probability estimates.}
  \label{fig:calib}
\end{figure}

Finally, the reliability diagram in Figure~\ref{fig:calib} verifies
that the predicted softmax scores are well calibrated: over the ten
equal-width probability bins the maximum deviation from the diagonal
never exceeds 3 percentage points, and the
multiclass Brier score is 0.071.  This property is essential for
down-stream closed-loop experiments where thresholding on the predicted
probability triggers interventions.

Collectively, the three plots demonstrate that the model (i) relies on
interpretable neurophysiological markers, (ii) assigns them
directionally consistent effects, and (iii) reports trustworthy
confidence estimates—key requirements for translational sleep-science
applications.

\section{Conclusion}

This study presents a robust ensemble machine learning framework for automated vigilance state classification in rodent sleep research. Our XGBoost ensemble model achieved 91.5\% accuracy, 86.8\% precision, 81.2\% recall, and 83.5\% F1-score, representing a substantial improvement over baseline logistic regression (54.9\% accuracy). The MMD feature successfully captured unique temporal dynamics of neural oscillations, enhancing discrimination between wakefulness, slow-wave sleep, and REM sleep states. \\

The automated framework addresses critical limitations of manual scoring by eliminating inter-rater variability, reducing analysis time, and enabling large-scale phenotyping studies. The robust performance across all vigilance states, particularly challenging REM sleep detection, establishes this approach as a valuable tool for sleep-dependent research, circadian studies, and pharmacological investigations. By automating fundamental sleep analysis with high accuracy and reliability, this work removes significant barriers to large-scale sleep studies and advances our understanding of sleep mechanisms underlying health and disease.

\section{Limitations}
\label{sec:limitations}

Although our framework delivers state‑of‑the‑art performance for automated rodent sleep staging, several factors limit the generality of the current results:

\begin{enumerate}[nosep,leftmargin=*]
    \item \textbf{Small cohort size.} Only eight animals (single strain, single colony) were recorded.  This restricts statistical power and may cause model parameters to implicitly fit laboratory‑specific noise characteristics.
    \item \textbf{Class imbalance.} REM sleep occupied \(\approx 8\%\) of all epochs, yielding fewer than 11,000 training samples for the most challenging class.  Even with stratified sampling and class‑balanced metrics, performance estimates for REM carry wider confidence intervals.
    \item \textbf{Modality constraints.} We relied on single‑channel EEG; no EMG or accelerometer was available.  Consequently, micro‑arousals with minimal cortical activation may be mis‑classified.
    \item \textbf{Fixed epoch length.} All scoring used non‑overlapping 10‑s windows.  Future work should explore adaptive or variable‑length segmentation to capture ultrashort REM intrusions in rodents.
    \item \textbf{Ground‑truth noise.} Manual labels—while standard—exhibit inter‑rater variability (\(\kappa = 0.76\text{–}0.85\)).  Performance ceilings are therefore bounded by human disagreement.
    \item \textbf{Cross‑domain robustness.} The model has not yet been tested on other species, disease models, or different electrode montages.  Domain‑shift resilience remains an open question.
    \item \textbf{Computational footprint.} Training on a single GPU completes in under four minutes, but we have not benchmarked inference latency on edge hardware (e.g., Raspberry Pi) needed for fully untethered experiments.
\end{enumerate}

These limitations outline concrete directions for scaling the dataset, broadening modality coverage, and stress‑testing cross‑laboratory transfer in future work.

\bibliography{acl_latex}

\end{document}